\documentclass{svproc}
\usepackage{url}
\usepackage{biblatex}

\usepackage{amsmath}
\usepackage{qcircuit}
\usepackage{hyperref}
\usepackage{url}
\usepackage{epsfig,graphicx}
\usepackage{subcaption}
\usepackage{breakurl}

\usepackage{bm}
\usepackage{braket}
\usepackage{amsfonts}
\usepackage[T1]{fontenc}
\let\ltxcup\cup
\usepackage{mathabx}
\usepackage[margin=1in]{geometry}

\begin{document}
\mainmatter              
\title{Universality for Sets of Three-Valued Qubit Gates}
\titlerunning{Universality for Universal Sets of Qutrit Gates}  
%
\author{Carlos Efrain Quintero Narvaez\inst{1}}
\authorrunning{Efrain Quintero} 
%
\tocauthor{Efrain Quintero}
\institute{Universidad Nacional Autonoma de México,\\
\email{efrainq07@ciencias.unam.mx}}

\maketitle              

\begin{abstract}
How to find universal sets quantum gates (gates whose composition can approximate any other gate with arbitrary precision) is an important part of the development of quantum computation science that has been explored in the past three decades with success \cite{Barenco_1995}, \cite{Cafaro_2011}. However, there has been little work in extending this very same theory to a generalization of qubits known as \textit{qudits} \cite{Vlasov_2002} (quantum units of information analogous to qubits that use any number $d\in \mathbb{N}$ of basis states instead of $2$ as qubits do). In this paper we will first do a review of the theory behind some essential proofs of quantum gate universality for qubit gates. After that we will show a new way of extending those statements to arbitrary qutrit (qudits with $3$ basis states) gates in an analogous manner. We also mention how could this be extended to any \textit{qudit} gate.
\end{abstract}
\section{Introduction}
In this paper first we will review Barenco's work \cite{Barenco_1995} on quantum gate universality. After that we will present the concept of three-valued qubits or as we call them here \textit{qutrits}. This will in turn lead to a new analogous proof of universality for systems with three-valued qubits which can eventually be adapted in future work for arbitrary-valued qubits or \textit{qudits} (units that use any number $d\in \mathbb{N}$ of basis states instead of $2$ as qubits do). \par
 Before continuing any further let us give a brief overview of some standard concepts of quantum computing, starting by the qubit. A qubit is usually said to be the quantum analog of a classical bit. While a classical bit can have one of two values $\{0,1\}$ a qubit can be in a linear superposition of two states $\{\ket{0},\ket{1}\}$ of the form:
    \begin{equation}\label{qubit definition}
        \ket{\psi}=\alpha \ket{0}+\beta \ket{1}
    \end{equation} 
    with $\alpha,\beta \in \mathbb{C}$ such that $|\alpha|^2+|\beta|^2=1$ in such a way that when it is measured it collapses to a values with probability $|\alpha|^2$ for $\ket{0}$ and $|\beta|^2$ for $\ket{1}$. Then, if we set the following column vectors:
    \begin{align}
        \ket{0}=\begin{pmatrix}
        1\\
        0
        \end{pmatrix}\quad\ket{1}=\begin{pmatrix}
        0\\
        1
        \end{pmatrix}
    \end{align}
    we can write the qubits state from \autoref{qubit definition} as a column vector in the following manner:
    \begin{align}
        \ket{\psi} = \begin{pmatrix}
        \alpha\\
        \beta
        \end{pmatrix}
    \end{align}
    The space of all possible qubit states is a two dimensional Hilbert space $\mathcal{H}(2)\cong \mathbb{C}^2$. And when one has a system containing $n\in \mathbb{N}$ qubits it is described as the tensor product of $n$ copies of the $\mathcal{H}(2)$ space we just mentioned which gives us $\mathcal{H}(2)^{\otimes n} \cong \mathbb{C}^{2^n}$, which in turn is a $2^n$ dimensional Hilbert space.\par
    In order to manipulate the information held by qubits it is necessary to design algorithms with a set of operations to do on them. For this consider that the set of transformations that can be done to these qubit states are the isometries of their respective Hilbert space. These transformations are known as the $N$ dimensional unitary group $\mathrm{U}(N)$ where each of its elements can be represented by an $N\times N$ unitary matrix with complex entries, where $N\in \mathbb{N}$ is the dimension of the corresponding Hilbert space that describes our quantum system. In the context of quantum computing these transformations (in this case corresponding to $\mathrm{U}(2^n)$ where $n$ is again the number of qubits) are referred to as quantum gates analogous to classic logic gates.\par
    Some examples of one qubit gates (elements of $\mathrm{U}(2)$) are the following:
    \begin{itemize}
        \item The $\mathrm{NOT}$ gate also known as the $X$ Pauli matrix. 
            \begin{equation}
            \label{equation:xpauli}
                X = \begin{pmatrix}
                0 && 1\\
                1 && 0
                \end{pmatrix}
            \end{equation}
        \item The Pauli matrices $X,Y$ and $Z$ where $X$ is exactly the $\mathrm{NOT}$ gate we just mentioned and the other two are as follows (where $i$ is the complex unit such that $i^2=-1$):
        \begin{align}
            Y &= \begin{pmatrix}
            0 && -i\\
            i&& 0
            \end{pmatrix}\\
            Z &= \begin{pmatrix}
            1 && 0\\
            0 && -1
            \end{pmatrix}
        \end{align}
        \item The Hadamard gate known for transforming the basis states $\{\ket{0},\ket{1}\}$ into uniform superpositions of those same states:
        \begin{equation}
        \label{equation:hadamard_gate}
            H = \begin{pmatrix}
            1/\sqrt{2} && 1/\sqrt{2}\\
            1/\sqrt{2} && -1/\sqrt{2}
            \end{pmatrix}
        \end{equation}
    \end{itemize}
    These gates are often represented with quantum circuits very similar to those of classical logic. For example the $X$ gate is represented as follows in Circuit \eqref{circuit:simple}:
    \begin{align}
    \label{circuit:simple}
        \Qcircuit @C=3em @R=.7em {
            &\gate{X} &\qw\\
        }
    \end{align}
    Here, the wire represents the ``flow`` of a qubit through time from left to right, and eventually it has the $X$ gate applied to it and continues. If we wanted represent the composition of two gates for example $YX$ we would have as follows in Circuit \eqref{circuit:simple2}:
    \begin{align}
    \label{circuit:simple2}
        \Qcircuit @C=3em @R=.7em {
            &\gate{X} &\gate{Y} &\qw\\
        }
    \end{align}
    Here $X$ is applied first and then $Y$ as a qubit flows through the wire from the left side of the circuit to the right side.\par
    These gates we have presented have all acted on just one qubit each, never doing anything to two qubits at the same time. However, we can have gates in $\mathrm{U}(2^2)$ that act on a $2$-qubit space $\mathcal{H}(2)\otimes \mathcal{H}(2)$ (which is of dimension $2^2=4$). A relevant example of these gates is the $\mathrm{CNOT}$ gate which in matrix form is:
    \begin{align*}
        \mathrm{CNOT} = \begin{pmatrix}
        1 && 0 && 0 && 0\\
        0&& 1 && 0 && 0\\
        0&& 0&& 0 && 1\\
        0&& 0&& 1 && 0
        \end{pmatrix}
    \end{align*}
    And is represented as a circuit as follows in Circuit \eqref{circuit:controlled_not}:
    \begin{align}
    \label{circuit:controlled_not}
        \Qcircuit @C=3em @R=.7em {
            &\ctrl{1} &\qw\\
            &\targ &\qw
        }
    \end{align}
    Notice how in Circuit \eqref{circuit:controlled_not} there are two wires representing one qubit each. That gate acts only on the first qubit, the one represented by the bottom wire, and it does so depending on the value of the second one which is called the \textit{control} qubit. This concept can be extended to an arbitrary number of control qubits, for example consider Circuit \eqref{circuit:toffoli} (the one shown below) with two of those:
    \begin{align}
    \label{circuit:toffoli}
        \Qcircuit @C=3em @R=.7em {
        &\ctrl{1} &\qw\\
            &\ctrl{1} &\qw\\
            &\targ &\qw
        }
    \end{align}
    Then the corresponding matrix representation for Circuit \eqref{circuit:toffoli} would be:
    \begin{align*}
        T = \begin{pmatrix}
        1&& 0 && 0 && 0 &&  0&& 0 && 0 && 0 \\
        0&& 1 && 0 && 0 &&  0&& 0 && 0 && 0 \\
        0&& 0 && 1 && 0 &&  0&& 0 && 0 && 0 \\
        0&& 0 && 0 && 1 &&  0&& 0 && 0 && 0 \\
        0&& 0 && 0 && 0 && 1 && 0 && 0 && 0\\
        0&& 0 && 0 && 0 && 0&& 1 && 0 && 0\\
        0&& 0 && 0 && 0 && 0&& 0&& 0 && 1\\
        0&& 0 && 0 && 0 && 0&& 0&& 1 && 0
        \end{pmatrix}
    \end{align*}
    This kind of gates are called the Toffoli gates. These gates act on the target qubit differently according to the value of the control qubits. This idea can be generalized as follows:
\begin{definition}[controlled gate]
\label{definition:controlled_gate}
    Take $U\in \mathrm{U}(2)$ (a 1-qubit gate) with a matrix representation given by:
    \begin{equation}
    \label{equation:controlled gate}
        U = \begin{pmatrix}
         u_{00}&& u_{01}\\
         u_{10}&& u_{11}
        \end{pmatrix}
    \end{equation}
    Where $u_{00},u_{01},u_{10},u_{11}\in \mathbb{C}$. Then define $\Lambda_{1}(U)\in \mathrm{U}(2^2)$ (a 2-qubit gate) with a matrix representation given by:
    \begin{equation}
        \Lambda_{1}(U) = \begin{pmatrix}
        1 && 0 && 0 && 0\\
        0 && 1 && 0 && 0\\
        0 && 0 && u_{00} && u_{01}\\
        0 && 0 && u_{10} && u_{11}
        \end{pmatrix}_{2^2\times 2^2}
    \end{equation}
    This $2^2\times 2^2$ matrix $\Lambda_{1}(U)$ object is called  the \textit{controlled} $U$ gate. 
\end{definition}
    Consider a two qubit state given as follows (with $\ket{\psi_2}\in \mathcal{H}(2)$ and $\ket{\psi_1}\in \mathcal{H}(2)$):
    \begin{equation*}
        \ket{\psi_1}\otimes\ket{\psi_2} \in \mathcal{H}(2)\otimes \mathcal{H}(2)\\
    \end{equation*}
    also see that $\ket{\psi_1}=\alpha\ket{0}+\beta\ket{1}$ for some $\alpha,\beta \in \mathbb{C}$ such that $|\alpha|^2+|\beta|^2=1$ and thus have:
    \begin{gather*}
        \ket{\psi_1}\otimes\ket{\psi_2} = (\alpha\ket{0}+\beta\ket{1})\otimes \ket{\psi_2}\\
        = \alpha\ket{0}\otimes \ket{\psi_2}+\beta \ket{1}\otimes \ket{\psi_2}
    \end{gather*}
    Having this, observe that controlled gates $\Lambda_{1}(U)$ as in the definition that was just given above have the property of taking two qubit states like the we just showed and transforming it into:
    \begin{equation*}
        \alpha\ket{0}\otimes\ket{\psi_2}+\beta\ket{1}\otimes(U\ket{\psi_2})
    \end{equation*}
    This means $\Lambda_{1}(U)$ applies $U$ to the part of the superposition whose first qubit is $1$ and otherwise it applies the identity $\bm{I}_2$. We call that first qubit the \textit{control qubit}.\par
    Another way to write the matrix representation of $\Lambda_1(U)$ is using matrix blocks inside another matrix, by this we mean as follows (considering $\bm{I}_2$ to be the $2\times 2$ identity matrix and $\bm{0}_2$ the $2\times 2$ zero matrix):
    \begin{equation*}
        \Lambda_1(U)=\begin{pmatrix}
        \bm{I}_2 && \bm{0}_2\\
        \bm{0}_2 && U
        \end{pmatrix}
    \end{equation*}
    \begin{definition}[$m$-qubit controlled gate]
        Again, let $U\in \mathrm{U}(2)$ be as in \autoref{equation:controlled gate}. Then let $\Lambda_{m}(U) \in U(2^{m+1})$ for $m\in \mathbb{N}$ be such that its matrix representation is given as follows:
        \begin{equation}
            \Lambda_{m}(U) = 
            \begin{pmatrix}
            \bm{I}_2 && \bm{0}_2 && \hdots && \bm{0}_2\\
            \bm{0}_2 && \bm{I}_2 && \hdots && \bm{0}_2\\
            \vdots && \vdots && \ddots && \vdots \\
            \bm{0}_2 && \bm{0}_2 && \hdots && U
            \end{pmatrix}_{2^{m+1}\times 2^{m+1}}
        \end{equation}
        For example we have that $\Lambda_2(U)\in \mathrm{U}(2^3)$ is:
        \begin{equation}
            \Lambda_2(U) = 
            \begin{pmatrix}
            \bm{I}_2 && \bm{0}_2&& \bm{0}_2&& \bm{0}_2\\
            \bm{0}_2&&\bm{I}_2 && \bm{0}_2 && \bm{0}_2\\
            \bm{0}_2&&\bm{0}_2 && \bm{I}_2 && \bm{0}_2\\
            \bm{0}_2&&\bm{0}_2 && \bm{0}_2 && U
            \end{pmatrix}_{2^3\times 2^3} =
            \begin{pmatrix}
            1 && 0 && 0 && 0 && 0 && 0 && 0 && 0 \\
            0 && 1 && 0 && 0 && 0 && 0 && 0 && 0 \\ 
            0 && 0 &&1 && 0 && 0 && 0 && 0 && 0 \\
            0 && 0 &&0 && 1 && 0 && 0 && 0 && 0 \\
            0 && 0 &&0 && 0 && 1 && 0 && 0 && 0 \\
            0 && 0 &&0 && 0 && 0 && 1 && 0 && 0 \\
            0 && 0 &&0 && 0 && 0 && 0 && u_{00} && u_{01} \\
            0 && 0 &&0 && 0 && 0 && 0 && u_{10} && u_{11}
            \end{pmatrix}_{2^3\times 2^3}
        \end{equation}
        Also note that $\Lambda_0(U) = U$.\newline
        Refer to all these as the $m$-qubit controlled $U$ gates. 
    \end{definition}
    With this notation, the controlled quantum gates mentioned before, the $\mathrm{CNOT}$ and the Toffoli gates, can be respectively written as $\Lambda_1(X)$ and $\Lambda_m(X)$ (for any $m\in \mathbb{N}$), where we used the $X$ Pauli matrix as defined in \autoref{equation:xpauli}.
    \section{Embeddings of $\mathrm{U}(n)$ into $\mathrm{U}(N)$ (with $n\leq N$)}
    Before continuing any further let us consider a simple case in order to build intuition. Consider the case of the group of isometries of the $N$ dimensional Euclidean space $\mathrm{O}(N)$ for some $N\in \mathbb{N}$ i.e. the $N$ dimensional orthogonal group. Then let $N=3$ in order to consider the $3$ dimensional Euclidean space. See that each element $M$ of $\mathrm{O}(3)$ can be represented as $3\times 3$ matrix with entries over $\mathbb{R}$:
    \begin{equation}
        M = 
        \begin{pmatrix}
        m_{00} && m_{01} && m_{02}\\
        m_{10} && m_{11} && m_{12}\\
        m_{20} && m_{21} && m_{22}
        \end{pmatrix}
    \end{equation}
    Now that we have this way of representing these transformations on 3 dimensional Euclidean space, consider how a rotation $M_{xy}\in \mathrm{O}(3)$ on the $x-y$ plane looks:
    \begin{equation}
        M_{xy} = 
        \begin{pmatrix}
        \cos\theta && -\sin\theta && 0\\
        \sin\theta && \cos\theta && 0\\
        0 && 0 && 1
        \end{pmatrix}\quad\quad\quad\theta\in \mathbb{R}
    \end{equation}
    When this transformation is applied onto a vector with three entries $v\in\mathbb{R}^3$ it only affects the first two, the ones corresponding to $x$ and $y$. Now consider rotations $M_{yz}$ and $M_{xz}$ on the $y-z$ and $x-z$ planes, respectively:
    \begin{align}
        &M_{yz} = 
        \begin{pmatrix}
        1 && 0 && 0\\
        0 && \cos\alpha && -\sin\alpha\\
        0 && \sin\alpha && \cos\alpha
        \end{pmatrix}
        &M_{xz} = 
        \begin{pmatrix}
        \cos\beta && 0 && -\sin\beta\\
        0 && 1 && 0\\
        \sin\beta && 0 && \cos\beta
        \end{pmatrix}
        &\quad\quad\quad\alpha,\beta\in \mathbb{R}
    \end{align}
    Just like with $M_{xy}$ when these matrices are applied onto a vector with three entries $v\in\mathbb{R}^3$ they leave one entry untouched (the one not related to the corresponding plane). These matrices only do a rotation around an axis (the $z$, $x$ and $y$ axes respectively) which implies they only rotate a subspace of the complete space they operate on (the $x-y$, $y-z$, and $x-z$ planes respectively), and thus we call them \textit{subspace rotations}. For rotating these planes, notice that we take a two dimensional rotation matrix $M_2\in \mathrm{O}(2)$:
    \begin{equation}
    M_2 = 
    \begin{pmatrix}
    \cos\theta && -\sin\theta\\
    \sin\theta && \cos\theta
    \end{pmatrix}\quad\theta\in\mathbb{R}
    \end{equation}
    And then we arrange the entries of this $2\times 2$ matrix onto a $3\times 3$ matrix as shown before. In a way this embeds a transformation from $\mathrm{O}(2)$ onto a higher dimensional group like $\mathrm{O}(3)$.
    This arrangements are done in this way because of the way matrix multiplication is defined, if we arrange the entries of a matrix in a certain way we will only affect certain entries of the vectors on which they operate. As this property comes almost exclusively from matrix multiplication, we can extend this idea of a \textit{subspace rotation} from $\mathrm{O}(N)$ to $\mathrm{U}(N)$ (the Lie group of unitary transformations on an $N$ dimensional complex Hilbert space) which can also be represented by matrices. We will see how this property is used in the following subsections.
    
    \subsection{Embedding of $\mathrm{U}(2)$ into $\mathrm{U}(N)$}
        Now consider again $U\in \mathrm{U}(2)$ with entries defined as before and consider the symmetry group $\mathrm{U}(N)$ as defined before for $N\in \mathbb{N}$ (not necessarily $N=2^n$). We're going to embed $U$ into this last group. We can do this in many ways, one of these is the gate $T^N_{1,0}(U)\in \mathrm{U}(N)$ defined as follows:
        \begin{equation*}
            T^N_{1,0}(U) = 
            \begin{pmatrix}
                u_{00} && u_{01} && 0 &&  && 0\\
                u_{10} && u_{11} && 0 && \hdots && 0 \\
                0 && 0 && 1 &&  && 0\\
                && \vdots &&  && \ddots &&  \vdots\\
                0 && 0 && 0 && 0 && 1
            \end{pmatrix}_{N\times N}
        \end{equation*}
        When this operator is applied onto an $N$ dimensional quantum state (represented as a column vector with $N$ complex entries) it will leave all entries untouched except for the $0$-th and the $1$-th which will be transformed by $U$ as if they formed a two row vector. \par
        For example, when $N=3$ we have that for the same $U$ we had before:
        \begin{align}
        T^{3}_{1,0}(U)=\begin{pmatrix}
        u_{00} && u_{01} && 0\\
        u_{10} && u_{11} && 0\\
        0 && 0 && 1
        \end{pmatrix}
        \end{align}
        More specifically for $X$ defined as in \autoref{equation:xpauli} we have:
        \begin{align}
        T^{3}_{1,0}(X)=\begin{pmatrix}
        0 && 1 && 0\\
        1 && 0 && 0\\
        0 && 0 && 1
        \end{pmatrix}
        \end{align}
    We can also write this idea of only affecting certain entries as follows, showing another form of embedding $\mathrm{U}(2)$ into $\mathrm{U}(N)$ (for an arbitrary value of $N$):
        \begin{align}
        \label{label:subspace2}
            T^N_{p,q}(U) &= 
            \begin{pmatrix}
            T_{ij}    
            \end{pmatrix}_{N\times N}\\
            T_{ij} &= 
            \begin{cases}
            \delta_{ij} & i\not\in \{p,q\}\text{ or } j\not\in \{p,q\}\\
            u_{00} & i=j=q\\
            u_{01} & i=q,\:j=p\\
            u_{10} & i=p,\:j=q\\
            u_{11} & i=j=p\\
            \end{cases}\nonumber
        \end{align}
        Where $p>q$ (this is done to adopt a convention on the ordering of the indices, however in principle it is valid to have $p<q$) and $p,q\in \{0,\hdots,N-1\}$. This operator will act on an $N$ dimensional quantum state vector $\ket{\psi_N} \in\mathbb{C}^N$ represented by a column vector as follows:
        \begin{align}
            \ket{\psi_N} = 
            \begin{pmatrix}
            c_0\\
            c_1\\
            \vdots\\
            c_{N-1}
            \end{pmatrix} =c_0\ket{0}+c_1\ket{1}+\hdots+c_{N-1}\ket{N-1}
        \end{align}
        Where $\ket{0},\ket{1},\hdots,\ket{N-1}$ are the canonical basis for $\mathcal{H}(N)$, i.e. the column vectors with zeros everywhere except on one entry where it has a $1$. \par
        This is important because $T^N_{p,q}(U)$ has the important property that it leaves every entry in a vector $\ket{\psi_N}$ untouched except for the $p$-th and $q$-th, which will have $U$ applied to them. This happens because this operator has exactly the same components as the identity matrix except for the ones that affect $\ket{p}$ and $\ket{q}$. This does a "complex rotation" on the $2$-dimensional subspace generated by $\ket{p}$ and $\ket{q}$.\par
\begin{figure}
            \centering
            \begin{subfigure}[b]{.45\linewidth}
                \includegraphics[width=\linewidth]{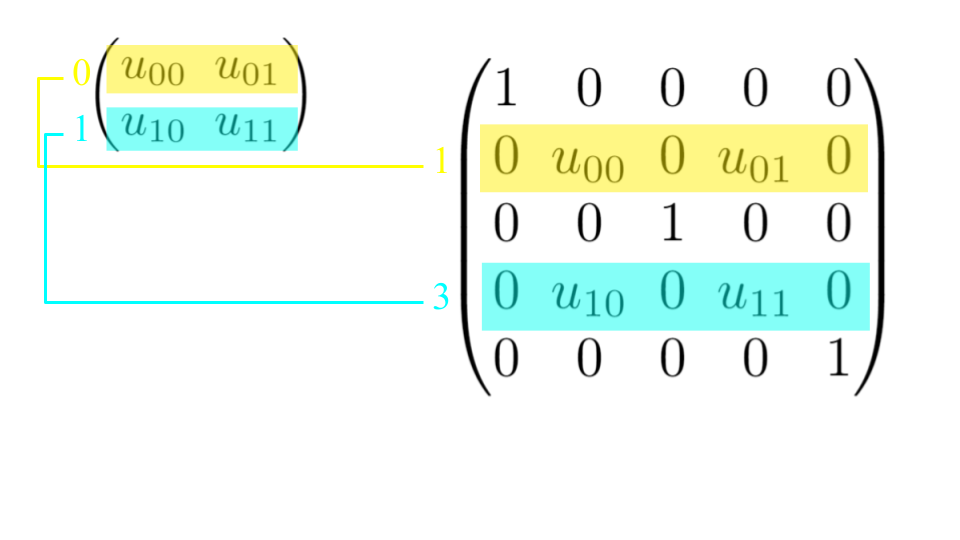}
                \caption{Identify rows.}
            \end{subfigure}
            \begin{subfigure}[b]{.45\linewidth}
                \includegraphics[width=\linewidth]{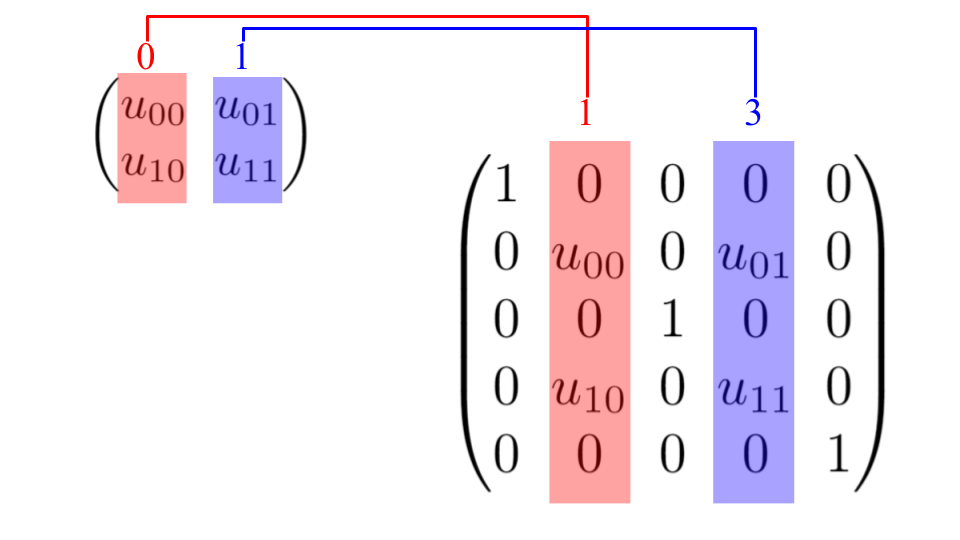}
                \caption{Identify columns.}
            \end{subfigure}
            \begin{subfigure}[b]{.45\linewidth}
                \includegraphics[width=\linewidth]{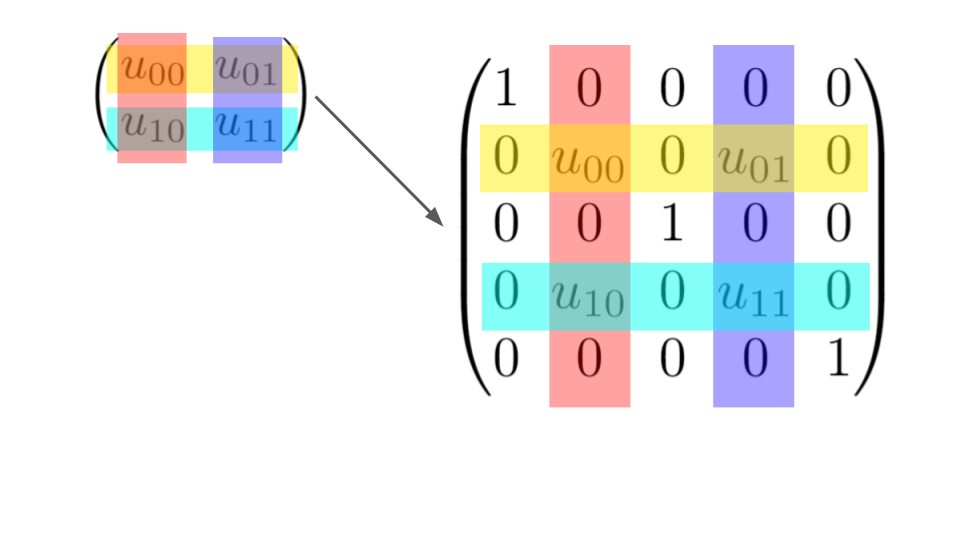}
                \caption{Result.}
            \end{subfigure}
            
            \caption{Intuition behind how the matrix $U\in \mathrm{U}(2)$ is turned into the matrix $T^5_{3,1}(U)\in \mathrm{U}(5)$ shown in \autoref{equation:higher_example}.}
            \label{figure:2d_subspace_rotation}
        \end{figure}

        Consider again the example with $N=3$, and let $p=2$ and $q=0$ then we get (using the same $U$ as before):
        \begin{align}
            T^{3}_{2,0}(U) = 
            \begin{pmatrix}
            u_{00}&&0&&u_{01}\\
            0&&1&&0\\
            u_{10}&&0&&u_{11}
            \end{pmatrix}
        \end{align}
        Accordingly consider then:
        \begin{align}
            T^{3}_{2,1}(U) = 
            \begin{pmatrix}
            1&&0&&0\\
            0&&u_{00}&&u_{01}\\
            0&&u_{10}&&u_{11}
            \end{pmatrix}
        \end{align}
        And lastly consider doing this embedding with the Hadamard gate defined on \autoref{equation:hadamard_gate}:
        \begin{align}
            T^{3}_{1,0}(H) &=
            \begin{pmatrix}
            1/\sqrt{2}&&1/\sqrt{2}&&0\\
            1/\sqrt{2}&&-1/\sqrt{2}&&0\\
            0&&0&&1
            \end{pmatrix}\quad
            T^3_{2,1}(H) &= \begin{pmatrix}
            1&&0&&0\\
            0&&1/\sqrt{2}&&1/\sqrt{2}\\
            0&&1/\sqrt{2}&&-1/\sqrt{2}\\
            \end{pmatrix}\quad
            T^3_{2,0}(H) &= \begin{pmatrix}
            1/\sqrt{2} && 0 && 1/\sqrt{2}\\
            0 && 1 && 0\\
            1/\sqrt{2} && 0 && -1/\sqrt{2}
            \end{pmatrix}
        \end{align}
        A higher dimensional example with $N=5$ is given as follows (using the same $U$ as before):
        \begin{align}\label{equation:higher_example}
            T^5_{3,1}(U) =
            \begin{pmatrix}
                1 && 0 && 0 && 0 && 0\\
                0 && u_{00} && 0 && u_{01} && 0\\
                0 && 0 && 1 && 0 && 0\\
                0 && u_{10} && 0 && u_{11} && 0\\
                0 && 0 && 0 && 0 && 1\\
            \end{pmatrix}
        \end{align}
        The intuition behind how to generate $T^5_{3,1}(U)$ is illustrated on \autoref{figure:2d_subspace_rotation}.
        \subsection{Embedding of $\mathrm{U}(3)$ into $\mathrm{U}(N)$}
        As we just did in the past subsection where we embedded a gate $U\in\mathrm{U}(2)$ onto a higher dimensional $\mathrm{U}(N)$, we can also do a similar embedding with an operator $V\in \mathrm{U}(3)$ with a matrix representation:
        \begin{align*}
            V = 
            \begin{pmatrix}
            v_{00} && v_{01} && v_{02}\\
            v_{10} && v_{11} && v_{12}\\
            v_{20} && v_{21} && v_{22}
            \end{pmatrix}
        \end{align*}
        where $v_{00}, v_{01}, v_{02},v_{10}, v_{11}, v_{12}, v_{20}, v_{21}, v_{22} \in \mathbb{C}$ and embed it into $\mathrm{U}(N)$ as we did before now with the gate $T^N_{p,q,r}(V)\in \mathrm{U}(N)$ (with $p,q,r\in \mathbb{N}$ and $p>q>r$) defined as follows:
        \begin{align}
        \label{label:subspace3}
            T^N_{p,q,r}(V) &= 
            \begin{pmatrix}
            T_{ij}    
            \end{pmatrix}\\
            T_{ij} &= 
            \begin{cases}
            \delta_{ij} & i\not\in \{p,q,r\}\text{ or } j\not\in \{p,q,r\}\\
            v_{00} & i=j=p\\
            v_{01} & i=p,\:j=q\\
            v_{02} & i=p, \: j=r\\
            v_{10} & i=q,\:j=p\\
            v_{11} & i=j=q\\
            v_{12} & i=q, \: j=r\\
            v_{20} & i=r,\:j=p\\
            v_{21} & i=r,\:j=q\\
            v_{22} & i=r \: j=r\\
            \end{cases}
        \end{align}
        Effectively creating an operator that leaves all entries of state vectors untouched except for the $p$-th, $q$-th and $r$-th entries to which it applies $V$. This again does a "complex rotation" to the $3$-dimensional subspace generated by $\ket{p},\ket{q}$ and $\ket{r}$.\par
\begin{figure}
            \centering
            \begin{subfigure}[b]{.45\linewidth}
                \includegraphics[width=\linewidth]{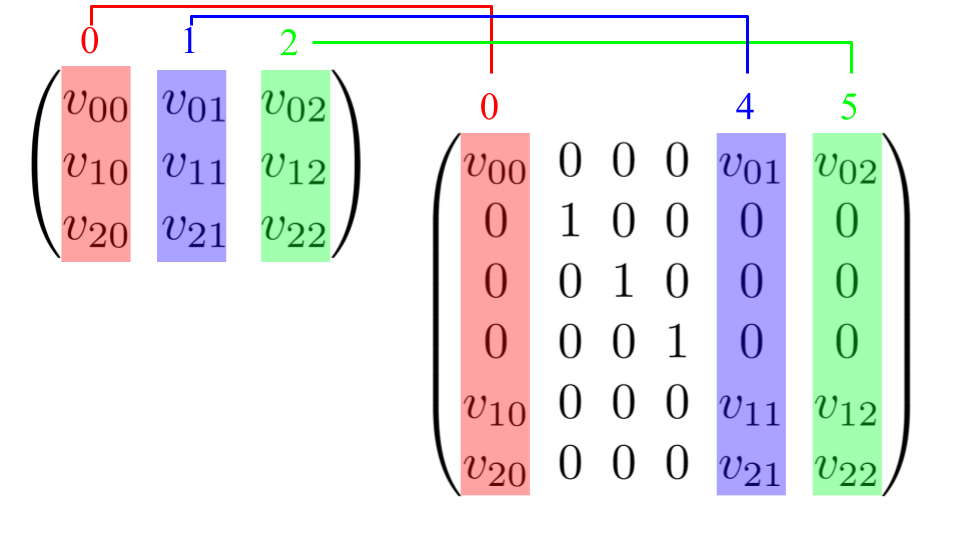}
                \caption{Identify rows.}
            \end{subfigure}
            \begin{subfigure}[b]{.45\linewidth}
                \includegraphics[width=\linewidth]{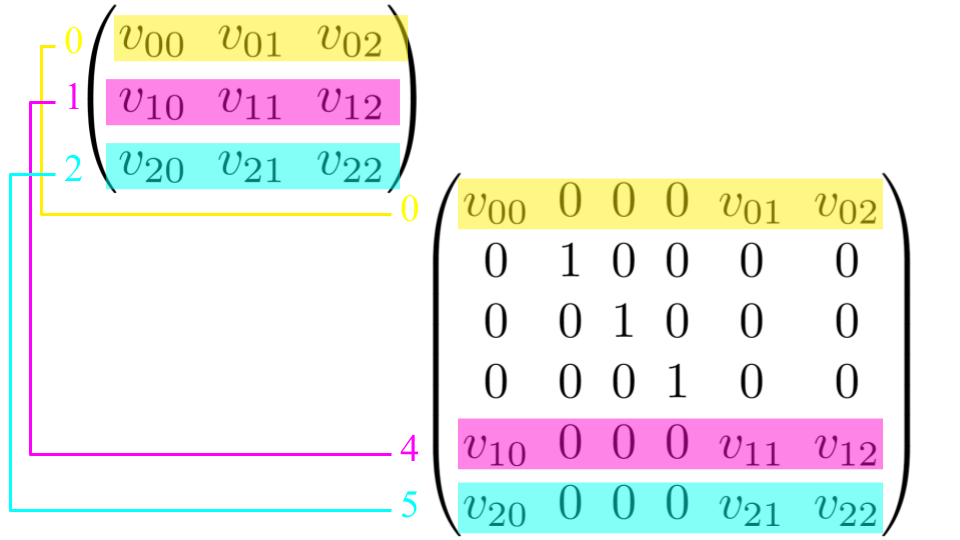}
                \caption{Identify columns.}
            \end{subfigure}
            \begin{subfigure}[b]{.45\linewidth}
                \includegraphics[width=\linewidth]{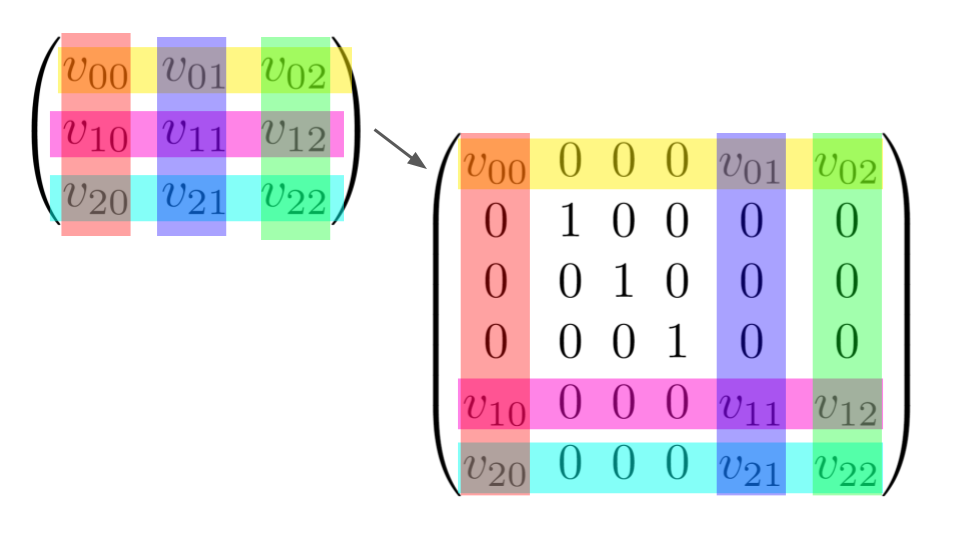}
                \caption{Result.}
            \end{subfigure}
            
            \caption{Intuition behind how the matrix $V\in \mathrm{U}(3)$ is turned into the matrix $T^5_{5,4,0}(V)\in \mathrm{U}(5)$ shown in \autoref{equation:even_higher_example}.}
            \label{figure:3d_subspace_rotation}
        \end{figure}

        We show some examples for the case when $N=6$, using the same $V$ matrix as before:
        \begin{align}
        \label{equation:even_higher_example}
            &T^6_{2,1,0}(V) = 
            \begin{pmatrix}
            v_{00} && v_{01} && v_{02} && 0 && 0 && 0\\
            v_{10} && v_{11} && v_{12} && 0 && 0 && 0\\
            v_{20} && v_{21} && v_{22} && 0 && 0 && 0\\
            0 && 0 && 0 && 1 && 0 && 0\\
            0 && 0 && 0 && 0 && 1 && 0\\
            0 && 0 && 0 && 0 && 0 && 1\\
            \end{pmatrix}
             &T^6_{4,2,0}(V) = 
            \begin{pmatrix}
            v_{00} && 0 && v_{01} && 0 && v_{02} && 0\\
            0 && 1 && 0 && 0 && 0 && 0\\
            v_{10} && 0 && v_{11} && 0 && v_{12} && 0\\
            0 && 0 && 0 && 1 && 0 && 0\\
            v_{20} && 0 && v_{21} && 0 && v_{22} && 0\\
            0 && 0 && 0 && 0 && 0 && 1\\
            \end{pmatrix}\\
            &T^6_{5,4,3}(V) = 
            \begin{pmatrix}
            1 && 0 && 0 && 0 && 0 && 0\\
            0 && 1 && 0 && 0 && 0 && 0\\
            0 && 0 && 1 && 0 && 0 && 0\\
            0 && 0 && 0 && v_{00} && v_{01} && v_{02}\\
            0 && 0 && 0 && v_{10} && v_{11} && v_{12}\\
            0 && 0 && 0 && v_{20} && v_{21} && v_{22}\\
            \end{pmatrix}
            &T^6_{5,4,0}(V) = 
            \begin{pmatrix}
            v_{00} && 0 && 0 && 0 && v_{01} && v_{02}\\
            0 && 1 && 0 && 0 && 0 && 0\\
            0 && 0 && 1 && 0 && 0 && 0\\
            0 && 0 && 0 && 1 && 0 && 0\\
            v_{10} && 0 && 0 && 0 && v_{11} && v_{12}\\
            v_{20} && 0 && 0 && 0 && v_{21} && v_{22}\\
            \end{pmatrix}
            \nonumber
        \end{align}
        The intuition behind the procedure for building these examples is illustrated in \autoref{figure:3d_subspace_rotation}.
        \subsection{Embedding of any $\mathrm{U}(n)$ into $\mathrm{U}(N)$ (with $n\leq N$)}
        We can define analogous transformations for any element of a $\mathrm{U}(n)$ embedded into a $\mathrm{U}(N)$ with $n\leq N$ and $n\in \mathbb{N}$. We call these subspace rotations.
        
    \begin{definition}[subspace rotation]
    Let us have $\mathrm{U}(N)$ with $N\in \mathbb{N}$ and consider $n\in \mathbb{N}$ such that $n\leq N$. Then consider $U\in \mathrm{U}(n)$ and define $T^{N}_{p_n,\hdots,p_1}(U) \in \mathrm{U}(N)$ (with $p_1,\hdots,p_n \in \{0,\hdots, N-1\}$ and $p_i > p_j$ if $i>j$ for any suitable $i,j\in \mathbb{N}$) to be the operator that does the complex rotation given by $U$ on the subspace generated by $\ket{p_1},\hdots, \ket{p_n}$ and leaves everything else unchanged as defined on equations \eqref{label:subspace2} and \eqref{label:subspace3}.\par As a side note, when it is understood from the context in which $\mathrm{U}(N)$ the subspace rotation is being done we will omit $N$ and just write $T_{p_n,\hdots,p_1}(U)$.
    \end{definition}
    Note that $\Lambda_m(U) = T^N_{N-1,N-2}(U)$ when $N=2^{m+1}$, that is to say \textit{controlled gates are subspace rotations}. This can be seen as we have that:
    \begin{equation}
            T_{N-1,N-2}^N(U) = 
            \begin{pmatrix}
            \bm{I}_2 && \bm{0}_2 && \hdots && \bm{0}_2\\
            \bm{0}_2 && \bm{I}_2 && \hdots && \bm{0}_2\\
            \vdots && \vdots && \ddots && \vdots \\
            \bm{0}_2 && \bm{0}_2 && \hdots && U
            \end{pmatrix}_{2^{m+1}\times 2^{m+1}}
        \end{equation}
    Which is exactly how the controlled gate $\Lambda_{m}(U)$ was defined.
    We'll use these definitions and properties in the following discussion of universality.
    
\subsection{Reck's decomposition for elements of $\mathrm{U}$(N)}
\label{section:reck_decomposition}
The discussion we just had introduced us to \textit{subspace rotations}, a method for embedding unitary transformations from some $\mathrm{U}(n)$ into a higher dimensional $\mathrm{U}(N)$ ($n\leq N$). We will use this notation along with the intuition we developed to present a result obtained by Reck on his 1994 article \cite{Reck_1994}. This result will in turn allow us to present Barenco's proof of universality for qubit gates \cite{Barenco_1995} later on.\par
First, consider an arbitrary unitary gate $M\in \mathrm{U}(N)$ for some $N\in \mathbb{N}$. As mentioned in Reck's paper we can always decompose the gate $M$ into several two-dimensional subspace rotations and a phase shift $\Phi$ as follows:
    \begin{align}\label{equation:reck}
        M &= \left(\displaystyle\prod_{i,j=0,\:j<i}^{N-1} R_{i,j}\right)\Phi
    \end{align}
    Where we have that $R_{i,j}\in \mathrm{SU}(N)$ and each of those is such that:
    \begin{equation}\label{equation:uij}
        R_{i,j} = T_{i,j}(U_{ij})
    \end{equation}
    for some $U_{i,j}\in \mathrm{SU}(2)$ i.e. a $2\times 2$ unitary matrix with unit determinant. The phase shift $\Phi$ is an $N\times N$ matrix with only elements of the form $e^{i\phi_k}$ for $k\in \{0,\hdots,N-1\}$ on its diagonal, i.e.:
    \begin{align}
        \Phi = \begin{pmatrix}
        e^{i\phi_0}&& && && &&\\
        && e^{i\phi_1} && && &&\\
        && && \ddots && &&\\
        && && && e^{i\phi_{N-1}}
        \end{pmatrix}_{N\times N}
    \end{align}
    Take for example $N=3$, then the decomposition of some $V\in \mathrm{U}(3)$ would be of the form:
    \begin{align}
        V &= R_{2,1}\: R_{2,0}\: R_{1,0}\: \Phi \label{equation:3_reck_example}\\
        &= T_{2,1}(A)\: T_{2,0}(B)\: T_{1,0}(C)\:\Phi
    \end{align}
    For some $A,B,C\in \mathrm{SU}(2)$ that take the same role that the $U_{ij}$ matrices took in \autoref{equation:uij}. Expanding all terms into their respective matrices we get:
    \begin{align}
        V = \begin{pmatrix}
        1 && 0 && 0\\
        0 && A_{00} && A_{01}\\
        0 && A_{10} && A_{11}
        \end{pmatrix}
        \begin{pmatrix}
        B_{00} && 0 && B_{01}\\
        0 && 1 && 0\\
        B_{10} && 0 && B_{11}
        \end{pmatrix}
        \begin{pmatrix}
        C_{00} && C_{01}&& 0\\
        C_{10} && C_{11} && 0\\
        0 && 0 && 1
        \end{pmatrix}
        \begin{pmatrix}
        e^{i\phi_0} && 0 && 0\\
        0 && e^{i\phi_1} && 0\\
        0 && 0 && e^{i\phi_2}
        \end{pmatrix}
    \end{align}
    Where $A_{ij},B_{ij},C_{ij}\in \mathbb{C}$ are the components of the $A,B,C$ matrices respectively.\par
    Now consider $N=4$ for another example, then we decompose $W\in \mathrm{U}(4)$ as follows (where we consider a different set of matrices $A,B,C$):
    \begin{align}
    W &= R_{3,2} \: R_{3,1}\: R_{3,0}\: R_{2,1}\: R_{2,0}\:R_{1,0} \: \Phi\\
    &= T_{3,2}(A) \:T_{3,1}(B) \:T_{3,0}(C) \:T_{2,1}(D) \:T_{2,0}(E) \:T_{1,0}(F)\: \Phi
    \end{align}
    For some $A,B,C,D,E,F \in \mathrm{SU}(2)$ that again take the some role as the $U_{ij}$ matrices took in \autoref{equation:uij}. Again, expanding all of these matrices we get (we have omitted the expansion of $\Phi$ for the sake of clarity):
    \begin{align}
        W = 
        \begin{pmatrix}
        1 && 0 && 0 && 0\\
        0 && 1 && 0 && 0\\
        0 && 0 && A_{00} && A_{01}\\
        0 && 0 && A_{10} && A_{11}\\
        \end{pmatrix}
        &\begin{pmatrix}
        1 && 0 && 0 && 0\\
        0 && B_{00} && 0 && B_{01}\\
        0 && 0 && 1 && 0\\
        0 && B_{10} && 0 && B_{11}\\
        \end{pmatrix}
        \begin{pmatrix}
        C_{00} && 0 && 0 && C_{01}\\
        0 && 1 && 0 && 0\\
        0 && 0 && 1 && 0\\
        C_{10} && 0 && 0 && C_{11}\\
        \end{pmatrix}
        \begin{pmatrix}
        1 && 0 && 0 && 0\\
        0 && D_{00} && D_{01} && 0\\
        0 && D_{10} && D_{11} && 0\\
        0 && 0 && 0 && 1\\
        \end{pmatrix}
        \begin{pmatrix}
        E_{00} && 0 && E_{01} && 0\\
        0 && 1 && 0 && 0\\
        E_{10} && 0 && E_{11} && 0\\
        0 && 0 && 0 && 1\\
        \end{pmatrix}
        \begin{pmatrix}
        F_{00} && F_{01} && 0 && 0\\
        F_{10} && F_{11} && 0 && 0\\
        0 && 0 && 1 && 0\\
        0 && 0 && 0 && 1\\
        \end{pmatrix}\Phi
    \end{align}
    Where we again consider the components of each of the $A,B,\hdots,F$ matrices to be represented by their respective symbols with the relevant subindices.\par We will show how this result is used in the context of universality of quantum gates in the following discussion.
    
\section{Universality for Qubit Gates}
Generally a set of qubit gates $\mathcal{S}$ (including a set of a single gate) is said to be universal if one can build any other gate for any number of qubits only with instances of the elements of that set. Another way of saying this is that a set $\mathcal{S}$ of quantum gates is universal if their composition can generate any gate in $U(2^n)$ for any $n\in \mathbb{N}$. This is valid as the set of gates for $n$ qubits is the set of unitary transformations on the state space of $\mathcal{H}(2)^{\otimes n}$, which is of dimension $2^n$. This notion of universality can sometimes be too strict for the purposes of experimental realization of an arbitrary gate. Often times one is not interested in effectively using a \textit{precise} quantum gate but an approximation of it. Thus there has been development into finding sets of gates that can \textit{approximate} any gate to a certain degree with a suitable number instances of gates from the set. A prominent example of such a set with \textit{approximate universality} is Boykin's universal set \cite{Boykin_2000}:
\begin{equation}
    \mathcal{S}^{\text{Boykin}}=\{H,Z^{1/4},\Lambda_1(X)\}
\end{equation}
Proving that this set is universal is not a trivial matter. On his article, Boykin cites an elementary universality proof mentioned by Barenco et al. \cite{Barenco_1995} which in turn facilitates the intuition of how any gate could in principle be formed by this set. This proof will be shown in the following discussion. 
    \subsection{Elementary Universality Proof}
    Barenco et al. in their 1995 paper \cite{Barenco_1995} present a proof of a very useful property for proving that certain sets of gates are universal. There, they consider specifically the set $\mathcal{S}=\{\mathrm{CNOT}\}\ltxcup\mathrm{SU}(2)$, and show that this set is universal as we'll see next.\par
    First of all, consider an arbitrary $n$-qubit gate $G\in \mathrm{U}(2^n)$. We'll show that there is a composition of elements of $\mathcal{S}$ that forms $G$. Now, as mentioned on \autoref{section:reck_decomposition} it is possible to decompose this matrix as in \autoref{equation:reck} (in this case the $N$ mentioned in \autoref{equation:reck} is such that $N=2^n$) and get:
    \begin{align}\label{equation:reck}
        G &= \left(\displaystyle\prod_{i,j=0,\:j<i}^{2^n-1} R_{i,j}\right)\Phi
    \end{align}
    Where again each $R_{i,j}\in \mathrm{U}(2^n)$ is defined to be:
    \begin{align}
        R_{i,j} = T_{i,j}(U_{i,j})
    \end{align}
    For some $U_{i,j}\in \mathrm{SU}(2)$. This means the gate $G$ can be obtained by composing a number of subspace rotations. Thus the problem is reduced to finding a way of building those subspace rotations with $\mathcal{S}$. For this, consider the following method:
    \begin{enumerate}
        \item Consider an arbitrary subspace rotation $T_{ij}(U)\in \mathrm{U}(2^n)$ for $U\in \mathrm{U}(2)$ and $i,j\in \mathbb{N}$, $0\leq i,j<2^{n}$.
        \item Write $i$ and $j$ in binary resulting in two $n$-bit strings, take for example $n=4$ and $i= 11$ and $j=4$, then we get:
        \begin{align*}
            i=11 &\longrightarrow 1011\\
            j=4 &\longrightarrow  0100
        \end{align*}
        \item Now consider the Grey code sequence for these two strings:
        \begin{align*}
            &1011\\
            &1010\\
            &1000\\
            &1100\\
            &0100
        \end{align*}
        This sequence gives you a way to transform the first string into the second doing one bit-flip at a time. This will be useful in the next steps.
        \item Now assume that we can generate any controlled gate using gates from $S$ (we'll show why this can be done later) and remember that any controlled gate is a two dimensional subspace rotation. For example see that $\Lambda_{n-1}(U) = T_{2^{n}-1,2^{n}-2}(U)$. However this only allows us to do rotations on the subspace generated by $\ket{2^{n}-2}$ and $\ket{2^{n}-1}$. This gives the idea of trying to find a way to change the entries on which the controlled gate acts. This will be done using Toffoli gates as follows (for the $4$-qubit case we presented earlier):
        \begin{align*}
        \Qcircuit @C=3em @R=.7em {
        &\quad\quad & \ctrl{3} & \qw\\
        &\quad\quad & \ctrl{2} & \qw\\
        & \quad\quad & \ctrl{1} &\qw\\
        &\quad\quad & \targ & \qw\\
        }
        \end{align*}
        The $4$-qubit gate represented in the circuit above checks whether the first three qubits are $1$ and only then does it apply the $\mathrm{NOT}$ or $X$ gate to the last qubit. This only affects the two basis states whose first $3$ qubits are all $1$'s. We can take advantage of this "checking property" to create a gate that only bit flips certain qubit. This is done as follows (using as an example the first two strings $1011$ and $1010$ of the Grey code sequence presented before):
        \begin{align*}
        \Qcircuit @C=3em @R=.7em {
        &\qw & \ctrl{3} & \qw & \qw\\
        &\gate{X} & \ctrl{2} & \gate{X} & \qw\\
        & \qw & \ctrl{1} &\qw & \qw\\
        &\qw & \targ & \qw & \qw\\
        }
        \end{align*}
        This gate checks whether the first three bits are exactly equal to $101$ (in order) and only then does it apply the $X$ gate to the last qubit. At the end it turns $1011$ into $1010$ which is what the Grey code requires on its first step. This is done by bit flipping the qubit we expect to be $0$ so that if it is indeed $0$ then after applying $X$ it becomes $1$ and makes the first three qubits be all $1$ when they reach the Toffoli gate, after that the qubit is bit flipped again to return its original state.\par
        Let's proceed with the next step of the Grey code (from $1010$ to $1000$), this is done with the following gate:
        \begin{align*}
        \Qcircuit @C=3em @R=.7em {
        &\qw & \ctrl{2} & \qw & \qw\\
        &\gate{X} & \ctrl{1} & \gate{X} & \qw\\
        & \qw & \targ &\qw & \qw\\
        &\gate{X} & \ctrl{-1} & \gate{X} & \qw\\
        }
        \end{align*}
        This gate now checks the first two qubits and the last one and only if the first two are equal to $10$ and the last one to $0$ does it apply the $X$ gate to the third qubit. This turns $1010$ into $1000$ as it is done in the Grey code.\par
        Lastly consider one last gate representing the second to last step of the Grey code (the actual last step isn't very useful as we'll see later):
        \begin{align*}
        \Qcircuit @C=3em @R=.7em {
        &\qw & \ctrl{1} & \qw & \qw\\
        &\qw & \targ & \qw & \qw\\
        & \gate{X} & \ctrl{-1} &\gate{X} & \qw\\
        &\gate{X} & \ctrl{-1} & \gate{X} & \qw\\
        }
        \end{align*}
        Now change all change all bits except the last one to be $1$ just for convenience in the next steps:
        \begin{align*}
        \Qcircuit @C=3em @R=.7em {
        & \qw & \qw\\
        & \qw & \qw\\
        & \gate{X} &\qw\\
        & \gate{X} &\qw\\
        }
        \end{align*}
        This way we have developed a method for generating a gate for each step of the Grey code sequence and having the following resulting values for $i$ and $j$:
        \begin{align*}
            i\longrightarrow 1111\\
            j\longrightarrow 0111
        \end{align*}
        \item Now compose the Grey code gates obtained before and get a gate that transforms $\ket{i}$ state into one that differs from $\ket{j}$ only in one bit on their bit string. We'll call this gate $K$. For our example we have the following resulting gate that transforms $1011$ into $1100$ which differs from $0100$ only in one bit:
        \begin{align*}
        \Qcircuit @C=3em @R=.7em {
        &\qw & \ctrl{3} &\qw & \ctrl{2} & \qw & \ctrl{1}&  \qw\\
        &\gate{X} & \ctrl{2} &\qw&  \ctrl{1} & \gate{X}  & \targ  & \qw\\
        & \qw & \ctrl{1} &\qw& \targ  & \gate{X} & \ctrl{-1}  & \qw\\
        &\qw & \targ & \gate{X} & \ctrl{-1} &\qw & \ctrl{-1}  & \qw\\
        }
        \end{align*}
        Note that the inverse of $K$ is obtained simply by reversing the order in which each of the Grey code gates is applied. For our example this results in:
        
        \begin{align*}
        \Qcircuit @C=3em @R=.7em {
        & \ctrl{1} & \qw & \ctrl{2} & \qw & \ctrl{3} & \qw & \qw\\
        & \targ & \gate{X} & \ctrl{1} & \qw & \ctrl{2} & \gate{X} &\qw\\
         & \ctrl{-1} & \gate{X} & \targ & \qw & \ctrl{1} &\qw &\qw\\
        & \ctrl{-1} & \qw & \ctrl{-1} & \gate{X} & \targ & \qw & \qw\\
        }
        \end{align*}
        \item After applying $K$, do the rotation $U$ on the two resulting strings that differ only by one bit and then reverse the bit flips using $K^{-1}$. The rotation is done simply by applying the $U$ gate to the qubit that differs for the two strings, but having it controlled so that it only acts on the two states we ended up with. In our example this is done as follows:
        \begin{align*}
        \Qcircuit @C=3em @R=.7em {
            & \multigate{3}{\:K\:} & \gate{U} &\multigate{3}{K^{-1}}&\qw\\
            &\ghost{\:K\:}  & \ctrl{-1} &\ghost{K^{-1}}&\qw\\
            &\ghost{\:K\:} & \ctrl{-2}  & \ghost{K^{-1}}&\qw\\
            &\ghost{\:K\:}  &\ctrl{-3} & \ghost{K^{-1}}&\qw\\
            }
        \end{align*}
        This resulting gate we built is $T_{ij}(U)$.
    \end{enumerate}
    
    With the procedure described before we can build any $T_{ij}(U)$ using only Toffoli gates, $X$ gates and $\Lambda_{N-1}(U)$. But remember our assumption that we could build the controlled gate for any $U$, this is done as follows:
    \begin{itemize}
        \item As proved by Barenco et al. in \cite{Barenco_1995}. For any $U\in \mathrm{U}(2)$ we can always find three operators $A,B,C \in \mathrm{SU}(2)$ such that:
        \begin{align}\label{equation:comp_prop}
            ABC &= \bm{I}_2\\
            A X B X C &= U
        \end{align}
        Where $X$ is the Pauli operator or the $\mathrm{NOT}$ gate.
        \item Having this then it is easy to see that we can build $\Lambda_1(U)$ as follows:
        \begin{equation}\label{equation:controlled_construction}
            \Qcircuit @C=3em @R=.7em {
                &\ctrl{1} &\qw\\
                &\gate{U}&\qw
            }
            =\Qcircuit @C=3em @R=.7em {
                &\qw    & \ctrl{1}  & \qw   & \ctrl{1}&\qw  &\qw\\
                &\gate{A}      & \targ     & \gate{B}   & \targ   &\gate{C}  &\qw
            }
        \end{equation}
    \end{itemize}
    Moreover it is indeed possible to prove that any Toffoli gate and $\Lambda_{n-1}(U)$ can be built using only $\mathrm{CNOT}$ and elements of $\mathrm{SU}(2)$ (as can be seen in \cite{Barenco_1995}). Knowing this and seeing that it is easy to build the phase shift $\Phi$ using only elements of $\mathrm{SU}(2)$ we can state that $\mathcal{S}$ is therefore a universal set of quantum gates.\par
    This last result is very useful when proving universality of a set as we now "only" need to prove that it can generate any gate in $\mathrm{SU}(2)$ and $\mathrm{CNOT}$.
    
    \section{Qutrits}
    Let us define precisely what we mean by a qutrit before proceeding any further. Similarly to how we defined a qubit, first consider a three valued classical unit of information which we'll call a \textit{trit} that can have one of three values $\{0,1,2\}$. Define then its quantum analog, a \textit{qutrit} which is a superposition of three values $\{\ket{0},\ket{1},\ket{2}\}$ as follows:
    \begin{equation}
        \ket{\psi}=\alpha \ket{0}+\beta \ket{1}+\gamma\ket{2}
    \end{equation} 
    with $|\alpha|^2+|\beta|^2+|\gamma|^2=1$. Notice then that we can denote this state as a column vector as we do with qubits:
    \begin{equation}
        \ket{\psi}=\begin{pmatrix}
        \alpha\\
        \beta\\
        \gamma
        \end{pmatrix}
    \end{equation} 
    Then the space of all qutrit states will be a three dimensional Hilbert space $\mathcal{H}(3)$. And when one has a system containing $n\in \mathbb{N}$ qutrits it is described again as the tensor product of $n$ copies of this Hilbert space $\mathcal{H}(3)^{\otimes n}$. This resulting Hilbert space is then of dimension $3^n$. The set of transformations or ``gates`` is then $\mathrm{U}(3)$ for a single qutrit and $\mathrm{U}(3^n)$ for $n$ qutrits.\par
    This allows us to have a three-valued logic approach to quantum computing that can be extended through the same method to have an arbitrary $n$-valued logic approach as we'll mention later.
    
    \section{Qutrit gates}
    Let us define some qutrit gates that will be used in the following discussion. First consider an analog of the $\mathrm{NOT}$ gate which we'll call the $\mathrm{TRANS}$ gate (short for \textit{transposition}):
    \begin{equation}
        \mathrm{TRANS}=
        \begin{pmatrix}
        1 && 0 && 0 \\
        0 && 0 && 1 \\
        0 && 1 && 0
        \end{pmatrix}
    \end{equation}
    This gate \textit{transposes} the last two entries of any qutrit state vector. Note that $\mathrm{TRANS}$ is a subspace rotation, specifically we have $\mathrm{TRANS} = T^{3}_{2,3}(X)$. We'll represent it in circuits as follows:\par
    \begin{equation}
        \Qcircuit @C=3em @R=.7em {
        &\gate{1,2} &\qw
        }
    \end{equation}
    The illustration symbolises the exchange of states $\ket{1}$ and $\ket{2}$. Note that we can also have a transposition operator that exchanges the other entries as follows:
    \begin{equation}
        \Qcircuit @C=3em @R=.7em {
        &\gate{1,2} &\qw\\
        &\gate{0,1} &\qw\\
        &\gate{0,2} &\qw
        }
    \end{equation}
    And we can add a subindex in the $\mathrm{TRANS}$ symbol to specify this and have:
    \begin{align}
        \mathrm{TRANS_{1,2}}&=
        \begin{pmatrix}
        1 && 0 && 0 \\
        0 && 0 && 1 \\
        0 && 1 && 0
        \end{pmatrix} = T^3_{1,2}(X)\\
        \mathrm{TRANS_{0,1}}&=
        \begin{pmatrix}
        0 && 1 && 0 \\
        1 && 0 && 0 \\
        0 && 0 && 1
        \end{pmatrix}= T^3_{0,1}(X)\\
        \mathrm{TRANS_{0,2}}&=
        \begin{pmatrix}
        0 && 0 && 1 \\
        0 && 1 && 0 \\
        1 && 0 && 0
        \end{pmatrix}= T^3_{0,2}(X)
    \end{align}
    However we'll omit this subindex and always refer to $T_{1,2}(X)$ when we mention $\mathrm{TRANS}$ as we can make the other transposition gates from this gate and one other as we will see next.\par
    Now consider the next gate which we'll call $\mathrm{ROT}$ (short for \textit{rotation}):
    \begin{equation}
        \mathrm{ROT}=
        \begin{pmatrix}
        0 && 1 && 0 \\
        0 && 0 && 1 \\
        1 && 0 && 0
        \end{pmatrix}
    \end{equation}
    If we consider this gate a permutation matrix then this kind of permutation is called a \textit{rotation}, hence the name. Its circuit representation is:
    \begin{equation}
         \Qcircuit @C=3em @R=.7em {
         &\gate{\circlearrowleft} &\qw\\
         }
    \end{equation}
    Also, notice that if we do three subsequent rotations we get the identity:
    \begin{equation}
        \mathrm{ROT}^3 = \bm{I}_3
    \end{equation}
    Now see the following:
    \begin{align*}
        (\mathrm{ROT})(\mathrm{TRANS}_{1,2})(\mathrm{ROT}^2)&=
        \begin{pmatrix}
        0 && 1 && 0 \\
        0 && 0 && 1 \\
        1 && 0 && 0
        \end{pmatrix}
        \begin{pmatrix}
        1 && 0 && 0 \\
        0 && 0 && 1 \\
        0 && 1 && 0
        \end{pmatrix}
        \begin{pmatrix}
        0 && 1 && 0 \\
        0 && 0 && 1 \\
        1 && 0 && 0
        \end{pmatrix}^2 =
        \begin{pmatrix}
        0 && 1 && 0 \\
        1 && 0 && 0 \\
        0 && 0 && 1
        \end{pmatrix} = \mathrm{TRANS}_{0,1}\\
        (\mathrm{ROT})(\mathrm{TRANS}_{0,1})(\mathrm{ROT}^2)&=
        \begin{pmatrix}
        0 && 1 && 0 \\
        0 && 0 && 1 \\
        1 && 0 && 0
        \end{pmatrix}
        \begin{pmatrix}
        0 && 1 && 0 \\
        1 && 0 && 0 \\
        0 && 0 && 1
        \end{pmatrix}
        \begin{pmatrix}
        0 && 1 && 0 \\
        0 && 0 && 1 \\
        1 && 0 && 0
        \end{pmatrix}^2 =
        \begin{pmatrix}
        0 && 0 && 1 \\
        0 && 1 && 0 \\
        1 && 0 && 0
        \end{pmatrix} = \mathrm{TRANS}_{0,2}
    \end{align*}
    Thus, we can obtain all the transposition gates from combining one of them with a number of rotation gates. This is why we defined $\mathrm{TRANS}$ the way we did, omitting indexes.\par
    Just like we did with the $\mathrm{NOT}$ and $\mathrm{CNOT}$ gates, we can add a control qutrit to the $\mathrm{TRANS}$ gate as follows:
    \begin{equation}
        \mathrm{CTRANS}=
        \begin{pmatrix}
        \bm{I}_{3} && \bm{0}_3 && \bm{0}_3\\
        \bm{0}_3 && \bm{I}_{3} && \bm{0}_3\\
        \bm{0}_3 && \bm{0}_3 && \mathrm{TRANS}
        \end{pmatrix}
    \end{equation}
    Where $\bm{I}_{3}$ and $\bm{0}_{3}$ are the $3\times 3$ identity and zero matrices respectively. This gate will only apply the $\mathrm{TRANS}$ gate to the target qutrit when the control one is in the state $\ket{2}$ otherwise it will do nothing. We represent this as circuit as follows:
    \begin{equation*}
        \Qcircuit @C=3em @R=.7em {
            &\ctrl{1} &\qw\\
            &\gate{1,2} &\qw\\
        }
    \end{equation*}
    Note that we can make the controlled versions of the other transposition gates as follows:
    \begin{equation*}
        \Qcircuit @C=3em @R=.7em {
            &\ctrl{1} &\qw\\
            &\gate{0,1} &\qw\\
        } 
        =
        \Qcircuit @C=3em @R=.7em {
            &\qw&\ctrl{1} &\qw&\qw&\qw\\
            &\gate{\circlearrowleft}&\gate{1,2} &\gate{\circlearrowleft}&\gate{\circlearrowleft}&\qw\\
        } 
    \end{equation*}
    \newline
    \newline
    \begin{equation*}
        \Qcircuit @C=3em @R=.7em {
            &\ctrl{1} &\qw\\
            &\gate{0,2} &\qw\\
        } 
        =
        \Qcircuit @C=3em @R=.7em {
            &\qw&\ctrl{1} &\qw&\qw&\qw\\
            &\gate{\circlearrowleft}&\gate{0,1} &\gate{\circlearrowleft}&\gate{\circlearrowleft}&\qw\\
        } 
    \end{equation*}
    Note that when the controlled gate on the right doesn't act, that is when it applies the identity operator on the lower qutrit, the result is a gate that applies the rotation gate three times which is of course the identity.
    \section{Universality for Qutrit Gates}
    Qutrit gate universality has the same corresponding definition as the one for qubit gates. A set $\mathcal{S}$ of qutrit gates is said to be universal if a combination of instances of gates of that set can build any gate for any number of qutrits. Another way of saying this is that for any $n\in \mathbb{N}$ any gate in $\mathrm{U}(3^n)$ (the set of unitary transformations in the space of qutrit states $\mathcal{H}(3)^{\otimes n}$) can be built by a composition of elements of $\mathcal{S}$.\par
    Having said this, similarly to what we reviewed before, we will present a new proof of universality for qutrit gates analogous to that of Barenco`s proof with qubits. Here we'll consider the set $\mathcal{S}= \{\mathrm{CTRANS}\}\ltxcup \mathrm{SU}(3)$ and show that it is universal as follows.\par
    Again we'll consider an arbitrary gate $G\in \mathrm{U}(3^n)$ for $n\in \mathbb{N}$, that is to say an $n$ qutrit gate. And again doing Reck's decomposition from \autoref{section:reck_decomposition}, specifically from equation \eqref{equation:reck} we get (in this case we have $N=3^n$):
    \begin{align}
        G &= \left(\displaystyle\prod_{i,j=0,\:j<i}^{3^{n}-1} R_{i,j}\right)\Phi
    \end{align}
    Where $R_{i,j}$ is defined as in \autoref{equation:uij} with (for some $U_{ij}\in \mathrm{SU}(2)$ same as before):
    \begin{align}
        R_{i,j} = T_{i,j}(U_{ij})
    \end{align}
    However, note this useful property of subspace rotations: if we consider $p,q,r\in \mathbb{N}$ with $0\leq p,q,r< 3^n-1$ then we have that there is some $J \in \mathrm{U}(3)$ such that:
    \begin{equation}
        R_{p,q}R_{q,r} = T^{3^n}_{p,q,r}(J)
    \end{equation}
    Thus we can write \eqref{equation:reck} as a composition of three dimensional subspace rotations by joining pairs as follows:
    \begin{equation}
        R_{p,q}R_{q,r} = R_{p,q,r}
    \end{equation}
    Where each $R_{p,q,r}$ is such that for some $U_{pqr}\in \mathrm{SU}(3)$ we have:
    \begin{equation}
        R_{p,q,r}=T^{3^n}_{p,q,r}(U_{pqr})
    \end{equation}
    Then write:
    \begin{align}
        G &= \left(\displaystyle\prod_{i,j,k=1,\:k<j<i}^{3^n-1} R_{i,j,k}\right)\Phi
    \end{align}
    Thus we have reduced the problem again to finding a way of building subspace rotations with $\mathcal{S}$, but in this case we have three dimensional subspace rotations.\par
    For this, consider the following analogous algorithm:
    \begin{enumerate}
        \item Consider an arbitrary $T^3_{i,j,k}(U)$ for $U \in \mathrm{U}(3)$ and $i,j,k\in \mathbb{N}$ with $0\leq i,j<3^n$.
        \item Write $i$,$j$ and $k$ in base 3 or \textit{ternary} resulting in three $N$ character strings, take for example $n=4$ and $i=52$, $j=45$ and $k=8$:
        \begin{align*}
            i=52\longrightarrow 1221\\
            j=45\longrightarrow 1200\\
            k=8\longrightarrow 0022
        \end{align*}
        \item Same as before, we'll make transformations on the base $3$ strings of $i,j,k$ until they have new values that only differ in one qutrit. For this, take the example given before.\par
        First step will be to change each of $i$,$j$ and $k$ first qutrits so they are $2$,$1$ and $0$ respectively. This will ensure that they differ in one qutrit, for this we'll apply the following gates:
        \begin{itemize}
            \item For changing $i=1221$ first qutrit from $1$ to $2$:
             \begin{align*}
                \Qcircuit @C=3em @R=.7em {
                &\gate{1,2}  & \ctrl{3} & \gate{1,2}  & \qw\\
                &\qw & \ctrl{2} & \qw & \qw\\
                & \qw & \ctrl{1} &\qw  & \qw\\
                &\qw & \gate{1,2} & \qw & \qw\\
                }
            \end{align*}
            \item For changing $j=1200$ first qutrit from $0$ to $1$:
            \begin{align*}
                \Qcircuit @C=3em @R=.7em {
                &\gate{1,2}  & \ctrl{3} & \gate{1,2}  & \qw\\
                &\qw & \ctrl{2} & \qw & \qw\\
                & \gate{0,2} & \ctrl{1} &\gate{0,2}  & \qw\\
                &\qw & \gate{0,1} & \qw & \qw\\
                }
            \end{align*}
            \item For changing $k=0022$ first qutrit from $2$ to $0$:
            \begin{align*}
                \Qcircuit @C=3em @R=.7em {
                &\gate{0,2}  & \ctrl{3} & \gate{0,2}  & \qw\\
                &\gate{0,2} & \ctrl{2} & \gate{0,2} & \qw\\
                & \qw & \ctrl{1} &\qw & \qw\\
                &\qw & \gate{0,2} & \qw & \qw\\
                }
            \end{align*}
        \end{itemize}
        This will result in now having the following values for $i$,$j$ and $k$:
        \begin{align*}
            i&\longrightarrow 1222\\
            j&\longrightarrow 1201\\
            k&\longrightarrow 0020
        \end{align*}
        We'll now do transformations character by character to make all of $j$'s characters be the same as those of $i$ except for the first one. We'll start by the leftmost character. However notice that the only one that differs in our examples if the second character so we only need to apply one gate:
            \begin{align*}
                \Qcircuit @C=3em @R=.7em {
                &\gate{1,2}  & \ctrl{1} & \gate{1,2}  & \qw\\
                &\qw & \ctrl{1} & \qw & \qw\\
                & \qw &  \gate{0,2} &\qw & \qw\\
                &\gate{1,2} & \ctrl{-1}&\gate{1,2} & \qw\\
                }
            \end{align*}
        Do the same transforming all of $i$'s characters except the first into those of $j$ applying the following gates:
        
        For changing $0020$ to $1020$:
             \begin{align*}
                \Qcircuit @C=3em @R=.7em {
                &\qw & \gate{0,1} & \qw & \qw\\
                &\gate{0,2} & \ctrl{-1} & \gate{0,2} & \qw\\
                & \qw &  \ctrl{-1} &\qw & \qw\\
                &\gate{0,2} & \ctrl{-1}&\gate{0,2} & \qw\\
                }
            \end{align*}
        For changing $1020$ to $1220$:
            \begin{align*}
                \Qcircuit @C=3em @R=.7em {
                &\gate{1,2}  & \ctrl{1} & \gate{1,2}  & \qw\\
                &\qw & \gate{0,2} & \qw& \qw\\
                & \qw &  \ctrl{-1} &\qw & \qw\\
                &\gate{0,2} & \ctrl{-1}&\gate{0,2} & \qw\\
                }
            \end{align*}
        This will give us values of $i,j,k$ that differ only in one character. In our example this will give us the following values:
        \begin{align*}
            i\longrightarrow 1222\\
            j\longrightarrow 1221\\
            k\longrightarrow 1220
        \end{align*}
        Finally, change all characters except the first to be $2$ just for convenience in the following steps:
        \begin{align*}
                \Qcircuit @C=3em @R=.7em {
                &\gate{1,2} &\qw\\
                &\qw &\qw\\
                &\qw &\qw\\
                &\qw &\qw
                }
            \end{align*}
        And we end up with:
        \begin{align*}
            i\longrightarrow 2222\\
            j\longrightarrow 2221\\
            k\longrightarrow 2220
        \end{align*}
        Compose all this gates in the order we gave them and call that $K$. As before, note that composing in reverse order gives the inverse $K^{-1}$.
        \item As done before use the controlled gate $\Lambda_{n-1}(U)$ (in this case $n-1=3$) to apply a subspace rotation to the transformed subspace which is now all represented in one qutrit, and form the following:
        \begin{align*}
        \Qcircuit @C=3em @R=.7em {
            & \multigate{3}{\:K\:}& \ctrl{1}& \multigate{3}{K^{-1}}&\qw\\
            &\ghost{\:K\:}&  \ctrl{1} &\ghost{K^{-1}}&\qw\\
            &\ghost{\:K\:}&  \ctrl{1}& \ghost{K^{-1}}&\qw\\
            &\ghost{\:K\:} &\gate{U} & \ghost{K^{-1}}&\qw\\
            }
        \end{align*}
        This will be the gate $T_{i,j,k}(U)$.
    \end{enumerate}
    Thus, we can build any gate $T_{i,j,k}(U)$ using $\Lambda_{n-1}(\mathrm{TRANS})$, $\mathrm{ROT}$, $\mathrm{TRANS}$ gates and $\Lambda_{n-1}(U)$. In order to construct these gates out of $\mathrm{CTRANS}$ and $\mathrm{SU}(3)$ as we said in the beginning notice the following property obtained when we apply \eqref{equation:reck} to a gate  $V\in \mathrm{U}(3)$ (as done in \autoref{equation:3_reck_example}):
    \begin{align}\label{equation:3_decomp_final}
        V &= R_{2,1}\: R_{2,0}\: R_{1,0}\: \Phi
    \end{align}
    Then, we can build any operator in $\mathrm{U}(3)$ out of 3 two dimensional subspace rotations. Moreover, one way to build $\Lambda_1(V)$ is by composing each of the controlled versions of these subspace rotations and the phase as follows:
    \begin{equation}
    \Qcircuit @C=3em @R=.7em {
    &\ctrl{1}&\qw\\
    &\gate{U}&\qw
    }
       = \Qcircuit @C=3em @R=.7em {
        &\ctrl{1}&\ctrl{1}&\ctrl{1}& \ctrl{1}&\qw\\
        &\gate{R_{2,1}}&\gate{R_{2,0}}&\gate{R_{1,0}}&\gate{\Phi}&\qw
        }
    \end{equation}
    Thus, if we can build the controlled version of any subspace rotation and phase change we'll be able to build any controlled qutrit gate. Now, notice the following property of subspace rotations for $V,W\in \mathrm{U}(2)$:
    \begin{equation}
        T_{i,j}(V)T_{i,j}(W) = T_{i,j}(VW)
    \end{equation}
    Then, for a subspace rotation $T^3_{i,j}(V)$ we have that as $V\in \mathrm{U}(2)$ then we can find $A,B,C \in \mathrm{SU(2)}$ that have the properties of \eqref{equation:comp_prop} and thus also do the following:
    \begin{align*}
        T_{i,j}(A)T_{i,j}(B)T_{i,j}(C) = T_{i,j}(\bm{I}_2)&= \bm{I}_{3}\\
        T_{i,j}(A)T_{i,j}(X)T_{i,j}(B)T_{i,j}(X)T_{i,j}(C) &= T_{i,j}(V)
    \end{align*}
    Then we can do a construction for $\Lambda_1(T_{i,j}(V))$ very similar to that of \eqref{equation:controlled_construction}, replacing $\mathrm{CNOT}$ with $\mathrm{CTRANS}$:
    \begin{align*}
        &\Qcircuit @C=3em @R=.7em {
        &\qw&\ctrl{1}&\qw&\ctrl{1}&\qw&\qw\\
        &\gate{T_{i,j}(A)}&\gate{i,j}&\gate{T_{i,j}(B)}&\gate{i,j}&\gate{T_{i,j}(C)}&\qw
        }
    \end{align*}
    Thus, as the subspace rotations used here are all in $\mathrm{SU}(3)$ (due to the fact that $A,B,C \in \mathrm{SU}(2)$) and we have seen how to create any controlled transposition gate using only $\mathrm{CTRANS}$ and $\mathrm{ROT}$ we find that we can build any $\Lambda_1(U)$ for $U\in \mathrm{U}(3)$.
    Using similar constructions as those of Barenco \cite{Barenco_1995} we can build any of the variants of $\Lambda_{n-1}(\mathrm{TRANS})$ and $\Lambda_{n-1}(U)$ using only $\mathrm{CTRANS}$, $\mathrm{ROT}$ and $\mathrm{SU(3)}$ (although $\mathrm{ROT}\in \mathrm{SU}(3)$). This is done through the same method we just used in \autoref{equation:3_decomp_final} where we divided $U$ into two dimensional subspace rotations and from there applied Barenco`s constructions exchanging $\mathrm{CNOT}$ with the appropiate variant of $\mathrm{CTRANS}$. And finally as done before we can see that it is easy to construct the $\Phi$ phase gate using $\mathrm{SU}(3)$.\par
    Thus, we have proved that $\{\mathrm{CTRANS}\}\ltxcup \mathrm{SU}(3)$ is an universal set for $\mathrm{U}(3^n)$. Again, as mentioned before in the qubit case, this makes it easier to prove that other sets are universal as we now ``only`` need to prove that they can form any gate in $\mathrm{SU}(3)$.
    \section{Conclusions}
    We reviewed some concepts of quantum computing including the notion of a universal set of gates. There we also reviewed the elementary proof of universality for qubit gates presented by Barenco et al. \cite{Barenco_1995}. After that we presented the concept of a qutrit and qutrit gates using a similar intuition to the one that was used in the discussion of qubits. Finally we adapted Barenco's proof for qubit gates to qutrit gates and gave a brief discussion of how this could be adapted to arbitrary valued qubits which we refer to as \textit{qudits}. This should allow us to speak more generally about universal sets of gates as this will no longer limit the use of such sets and formalism to the usual qubit states and transformations.\par
    \section{Acknowledgements}
    I would like to thank my advisor D. Berenice-Cabrera from UNAM in Mexico City for useful discussion regarding the content of this article as well as support for finding important citations for the ideas presented here. I would also like to express my deepest appreciation to my friends O. García-Esparza from ITESM in Monterrey, J. Rojas-Noble and M. Lopez-Valle from UNAM in Mexico City for useful remarks regarding the clarity and conciseness of this work. 
    \newpage
 \printbibliography
\end{document}